      [1999/12/01 v1.4c Il Nuovo Cimento]
\documentclass[varenna]{cimento}
\usepackage{graphicx}
\usepackage{multicol}

\title{Metallic conduction, apparent metal-insulator\\
transition and related phenomena  in two-dimensional electron
liquid}
\shorttitle{Metal-insulator transition in 2D}
\author{V.~M.~Pudalov
} \institute{P.~N.~Lebedev Physics Institute - 53 Leninskii
prospekt, Moscow, 119991 Russia}

\PACSes{\PACSit{71.30.+h, 73.40.Qv, 71.27.+a}
{
} }
\begin{document}

\maketitle

\begin{abstract}
The paper introduces a reader to
a relatively young
field of
the physics of strongly interacting and disordered
2D electron system, in particular, to the phenomena of the metallic conduction
and the apparent metal-insulator transition in 2D. The paper briefly overviews the
experimental data on the electron transport, magnetotransport, and spin-magnetization in 2D
and on the electron-electron interaction effects.
\end{abstract}

\section{Introduction}
Resistivity of materials around us ranges by more than 30 orders
of magnitude, from $10^{20}$ to $10^{-12}$ Ohm\,m. The variety of
materials are classified as Metals (M) and Insulators (I), which
are two distinct classes of matter only in the $T\rightarrow 0$
limit. The principle difference between the two classes is in the
character of the electronic wavefunctions, which are spatially
localized in insulators and extended in metals.

It is well-known,
that subjected to even insignificant changes of pressure,
dopant density,  etc.
many materials exhibit transformations between the  two states \cite{mott}.
The M-I transitions (MIT), unrelated to changes in the lattice structure and symmetry,
are of a special interest because they are considered to be
continuous Quantum phase transitions,  occurring
at $T=0$.
The M-I transitions usually take place as density
of electrons decreases;
the latter  necessarily
leads to the increase of the effective
strength of $e-e$ interaction.
A dimensionless  ratio $r_s$ of the potential (Coulomb) energy
to kinetic (Fermi)  energy is often used to quantify interaction
in 2D, $r_s \propto n^{-1/2}$ with $n$ being the electron density \cite{ando_review}.
On the other hand, the  decrease in the density is accompanied with
the increase in the effective disorder, particularly, due to the weakening of the
screening of potential fluctuations. Thus, both, interactions and disorder
are important in the vicinity of the MIT; the separation and interpretation
of their effects  represents a hard task \cite{Nagaoka_review}.

In the limit of strong interactions $r_s \rightarrow \infty$ and zero disorder,
the groundstate of the 2D system
is believed to be the Wigner crystal (WC) of electrons.
Correspondingly, at zero temperature and zero disorder, for 2D system, there is at least one
critical  point on the interaction axes,
quantum melting of the Wigner crystal  \cite{ceperley,ando_review}.
A weak disorder
is expected to  stabilize the WC  \cite{chui&tanatar,giamarchi}.
As disorder increases further, WC looses
long-range order and crosses over to a  localized
phase.
Thus, in the limit of
strong disorder and weak interactions, the ground state of the system
is built of the single-particle
localized states \cite{efros&shklovskyi}.
The  experiments always take place at non-zero disorder
and temperature.
 The subject of this paper
is related to phenomena which occur due to the combined action of
strong inter-electron interactions and disorder.

The outline of this paper is as follows:
Sections 1 and 2 introduce a reader to
the field and briefly review the main results in
the semiclassical and quantum frameworks, correspondingly.
Section 3 describes major results on transport in 2D systems, which are
discussed in terms of the  apparent metal-insulator  transition in 2D.
Section 4 describes quantitative studies of the effects of electron-electron interaction;
the experimental results are compared with available theories in various regimes.

\vspace{0.1in}
\underline{1.1. Mott semiclassical picture of the MIT}.\\
At finite temperature, or for finite system
size $L$, the metal and insulator states are not well-defined; one can be mislead
by attempting to distinguish these two states according to
their conduction. In fact, the two states are inherently related with  two
distinct types of electron wavefunctions, localized and extended ones.
The latter are classified according to the localization length $\xi$: if $\xi >
(r_2-r_1)$ the waves are extended (metal) and $\Psi \propto
\cos(kr)$; if $\xi < (r_2-r_1)$  the waves are localized
(insulator) and $\Psi \propto \exp(-r/\xi)\cos(kr)$. Note, the
electron spacing, ($r_2-r_1) \sim \lambda_F = 2\pi/k_F = 1120
\left(10^{11} g_v/2n \right)^{1/2}$\AA\ (the valley
degeneracy $g_v=2$  corresponds to (100)Si-crystal plane).

For the extended states, when $\lambda_F \ll l <L$ (with $l$ being the
mean free path) the electron motion  results in a
classical Drude conduction $\sigma = ne^2\tau_p/m=ne^2 l/v_Fm$.
As disorder increases, $l$ decreases. However, $l$ can not be
less than $\lambda_F$ (the so called Ioffe-Regel criterium
(1960)):
 $l_{\rm min}=\lambda_F = 2\pi/k_F$.
Thus, we obtain the minimum metallic conductivity \cite{mott}
$\sigma_{\rm min}(l_{\rm min})=2\pi e^2/h \quad {\rm in\ 2D}$.
We used here
$k_F^{2D}=\sqrt{2\pi n}$. Note, that $h/e^2=25812.7$\,Ohm.
In this semiclassical picture \cite{mott},
the M-I transition is caused by the disappearance
of the localized (or, vice versa, extended) states.

\vspace{0.1in}
\underline{1.2. Basics of the 2D semiconducting devices}\\
The 2D electron and hole systems discussed in this lecture consist of the bulk
carriers confined in a two-dimensional potential well; the latter is
 formed at the interface between two semiconductors
or semiconductor/insulator interface. Figure 2 schematically shows
crossview of the silicon metal-oxide-semiconductor (MOS) stucture \cite{ando_review}.
The 2D layer of electrons is confined at the interface between Si and SiO$_2$, grown at the
top of the Si-substrate. On the top of the insulating layer,
a thin metallic film (gate) is deposited.
Throughout this lecture, the electrons and holes are  the same quasiparticles
and for simplicity will be often called ``electrons''.

\begin{figure}[htb]
\begin{minipage}[t]{.47\textwidth}
\includegraphics[width=2.4in]{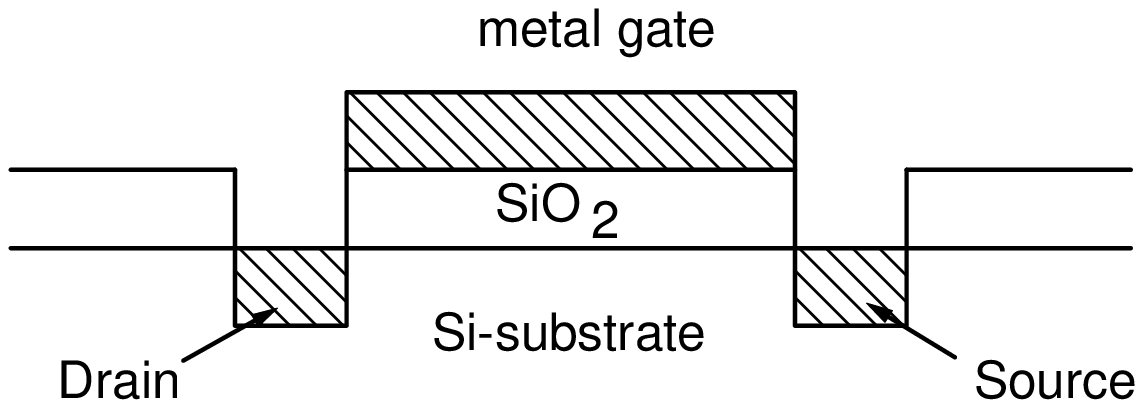}
\caption{Schematic structure of a silicon field effect transistor.
The 2D electron gas is
confined at the Si side of the  Si/SiO$_2$ interface.
Drain and source are diffusion areas
used as ohmic contacts to the 2D layer.
Metallic gate is deposited on the top of the SiO$_2$ layer
\protect\cite{ando_review}.}
\label{Si-MOSFET schematics}
\end{minipage}
\hfil
\begin{minipage}[t]{.47\textwidth}
\begin{center}
\includegraphics[width=1.9in]{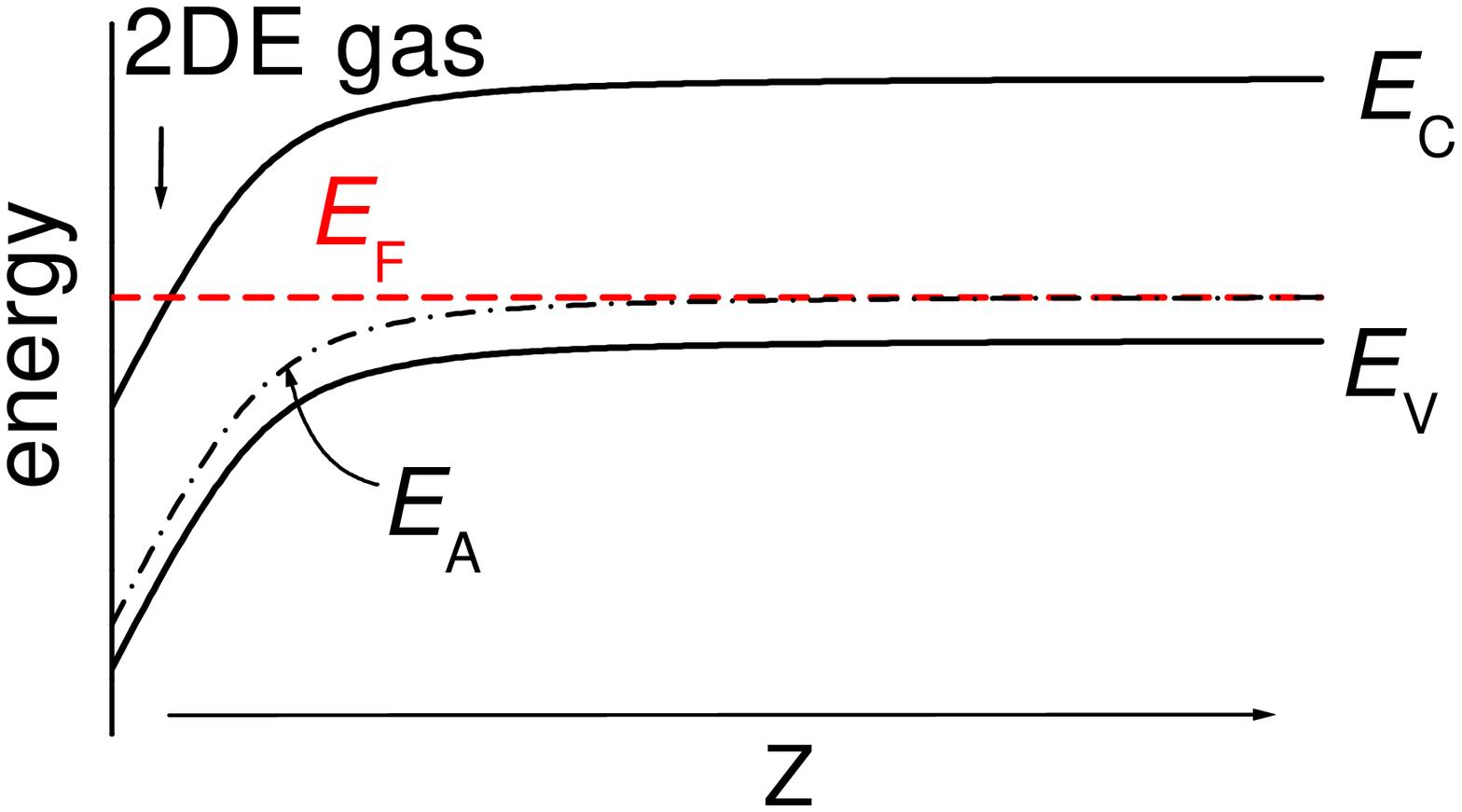}
\caption{Schematic energy band diagram of the Si-MOS structure. $E_C$ ($E_V$) is the bottom
(top) of the conduction (valence) band, $E_A$-bulk
acceptors energy level, $E_F$ - Fermi level,
$Z$-axes is directed to the bulk, perpendicular to the surface.}
\label{Energy band_Si-MOS}
\end{center}
\end{minipage}
\end{figure}

When the  positive voltage $V_g$ is applied to the gate (relative to
the source, or drain contact, or relative to the Si-bulk),
the conduction $E_c$ and valence $E_v$ band edges
bend down as Figure \ref{Energy band_Si-MOS} shows. For a sufficiently high gate voltage,
the bottom of the conduction band
decreases below the bulk Fermi level and the resulting triangular potential well
starts populating with electrons \cite{ando_review}. At room temperatures, when the bulk Si
is conducting, redistribution of the electrons between the bulk and the 2D
layer occurs in accord with the Poisson
equation \cite{ando_review}. When the gate voltage is applied or changed at low temperatures,
the bulk conduction is frozen-out, the equilibrium with  bulk is not achieved,
and the required electrons come into (or out) the 2D layer
from the potential contacts. The contacts are lithographically defined dopant diffusion areas
with high concentration of electrons supplied by the dopants.
\vspace{0.1in}
For our purpose, it is essential
only that the layer of electrons confined at the interface
is two-dimensional,
and that its charge $Q$ is controlled by voltage $V_g$
applied to the gate:
$Q\equiv ne = C(V_g-V_t)$, with $V_t$, the so called threshold voltage.
Thus, the Si-MOS system  may be viewed as
a plane capacitor,
whose one plate is formed by the 2D layer of electrons whereas the gate
serves as another plate.

\section{Quantum transport at zero field}
\underline{2.1. Various types of transport:
Delocalized states: diffusive and ballistic transport.}\\
There are two regimes of the metallic-like transport:
diffusive for $l < l_\varphi< L$, and ballistic regime for  $l> L$.
The phase breaking length $l_\varphi = \sqrt{D\tau_\varphi}$ (with $\tau_\varphi$,
the phase breaking time, $D$ - diffusion coefficient)
is related with
large changes of electron energy  and is similar (not equivalent though)
to inelastic length.
The relevant scattering processes for $\tau_\varphi$
at high temperatures are (i) electron-phonon scattering,
 and  for  low temperatures - (ii) electron-impurity  and
 (iii) electron-electron scattering.
 In general, $\tau_\varphi \propto T^{-p}$ with $p=1 - 2$, depending
 on the dominating scattering process.
Since electron-electron collisions conserve the total momentum,
the transport time $l_p$
(momentum relaxation)
is determined by the two former processes and is insensitive to
electron-electron scattering.

\vspace{0.1in}
\underline{2.2. Localized states}.\\
If an energy  barrier $\Delta$ separates the energy of the
electron states
from conduction (or valence) band, and the
wavefunctions of localized states
don't overlap, the conduction occurs
via temperature activated transitions whose probability is
$\sigma \propto \exp(-\Delta/T)$.
When the probability of the temperature activation
is too low (e.g., for high  $\Delta$,  or for low temperatures), transport occurs via
elastic tunneling between the localized states. This is the so-called
hopping conduction regime, where
the characteristic temperature dependence of the conductivity is
$\sigma \propto \exp((-T_0/T)^p)$,
where $p=1/2$ or 1/3, depending on a specific model \cite{efros&shklovskyi}.
Usually, hopping conduction 
dominates at low temperatures.

\begin{figure}
\begin{center}
\includegraphics[width=3.2in]{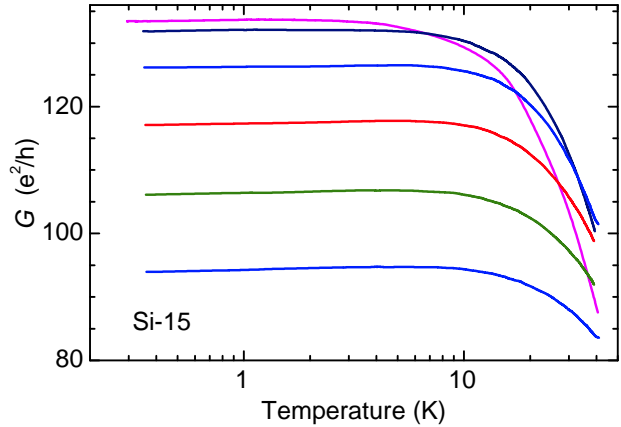}
\caption{$\sigma(T)$ dependences
for high densities \protect\cite{Gmax} (from top to bottom,
in units of $10^{11}$\,cm$^{-2}$):
32.1, 42.94,
48.4,  53.8,  64.7,
75.6,
86.5.}
\label{Gmax}
\end{center}
\end{figure}

\vspace{0.1in}
\underline{2.3. Electron's Phase
Coherence and Transport.}\\
In the quantum-mechanical picture, the electron waves propagate and interfere.
The interference gives
rise to  quantum corrections to the
conductivity \cite{altshuler&aronov,lee_review}.
In 2D system of non-interacting electrons (Fermi-gas), as $T\rightarrow 0$:
\begin{equation}
\sigma = \sigma_D
-\frac{e^2}{\pi h}\ln(\tau_{\varphi}/\tau) \sim
\sigma_D + \frac{e^2}{\pi h}\ln(T\tau),
\end{equation}
where $\sigma_D$ is the semiclassical Drude  conductivity.
The single-particle interference thus may be viewed as a quantum ``backscattering''.
Weak logarithmic dependence is a typical attribute of the 2D system at high electron
density (low $r_s$), and may be observed at high conductivity
$\sigma \gg e^2/h$ in the low-temperature
diffusive regime $T\tau \ll 1, \tau \ll \tau_\varphi$ (see fig.~\ref{Gmax}).

\vspace{0.1in}
\underline{2.4. Suppression of the weak localization in $H_\perp$ fields}\\
The quantum interference corrections originate from
small loops of self-intersections on the  trajectories, which quantum
particles  propagate in both directions, due to the  time reversal symmetry
\cite{altshuler&aronov,lee_review}.
The amplitudes of the wavefunction for a particle passing the loop clockwise and
counterclockwise acquires additional factors:
$
A_1 \rightarrow A_1\exp\left(\pm i\pi\Phi/\Phi_0\right)
$
with $\Phi_0=h/e$ being the flux quantum and $\Phi$ the magnetic flux through the loop.
The magnetic field applied perpendicular to the 2D plane of motion,
destroys the interference by reducing the probability for a particle to return.
The interference breaks down when the phase difference $\Delta\varphi$ becomes  $\sim 1$:
\begin{equation}
\Delta \varphi \sim \frac{H D \tau_\phi}{\Phi_0} \sim 1 \hspace{0.5in} H\sim \frac{h}{eD\tau_\varphi}
\end{equation}
The ``negative magnetoresistance'' effect is
a tool to determine $\tau_\varphi$ \cite{lee_review,fukuyama82}.

\vspace{0.1in}
\underline{2.5. Single-particle scaling theory of localization (E.\ Abrahams et al., 1979)
\cite{gang4}}\\
The one-parameter scaling theory considers
the dependence of the conductivity
on the system size.
When the size
$L$ is changed, the effective disorder is changed; it is
assumed that the only measure of this is the conductance $G$.
The above assumption is equivalent to
introducing a function $\beta$ such that
\begin{equation}
\beta = \frac{L}{G} \frac{dG}{dL}=\frac{d\ln G}{d\ln L},
\end{equation}
where $\beta$ is a universal function of $G$ solely.

\begin{figure}
\begin{center}
\includegraphics[width=2.4in]{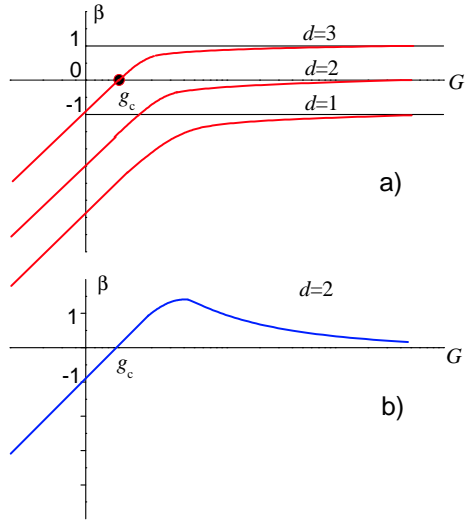}
\caption{a) Schematic behavior of the scaling function for
non-interacting electrons in different dimensions $d$. Bold dot depicts
the critical point of the M-I transition for  $d=3$.
b)Spin-orbit scattering case for $d=2$.} \label{beta}
\end{center}
\vspace{-0.2in}
\end{figure}

One can consider how the scaling hypothesis fits various obvious limits:\\
(1) In the Ohm's law  region ($G \gg e^2/h$),  $\beta= d-2$ and
the conductance
does not depend on $L$
for the two-dimensional case, $d=2$.\\
(2) If the states are localized,
$ G(L) \propto \exp(-L/\xi)$.
Therefore,
$\beta(G)=\ln(G/G_0)$. \\
The two distinct limits enable to plot the asymptotes of $\beta$ in fig.~\ref{beta}\,a.
One may expect that
between these two limits, $\beta $ should vary smoothly,
as shown schematically in fig.~\ref{beta}\,a.

\underline{Note:}
It is not necessary to vary $L$ in order to move along the scaling trajectories.
For finite temperatures, $l_\varphi$ causes  inelastic cut-off, so that the temperature
tunes the effective sample length.

In $d=3$, as fig.~\ref{beta}\,a shows,  the scaling function  $\beta <0$
for very small $G$ (strong disorder). As conductance increases, $\beta$
passes through zero at a certain critical value $G_c$.
This is a repulsive point or an ``unstable critical point'' in  the renorm-group
terminology.
For low initial $G<G_c$, increasing the size of the system  will cause $G$ to decrease
and the behavior of  the system approaches that for the localized states.
In contrast, when the initial conductance is high,
$G>G_c$, increasing the size of the system will leads to the Ohmic regime
(metallic conduction). Such behavior and the corresponding metal-insulator transition
are consistent with the Mott' semiclassical picture
described above.

In lower dimensions, $d<3$,
$\beta $ is always negative and all states are localized. As system size $L$ increases,
the system moves to the
strongly  localized regime. The scaling function describes a crossover from
weak localization (localization length
$\xi >L$) to strong localization ($\xi <L$) regime.

It follows from the
one-parameter scaling theory   that
``there is no true metallic conduction in $d<3$'' \cite{gang4}.
For $d=1$, this conclusion is obvious since the negative scaling function has
substantially large amplitude $|\beta| \geq 1$. In accordance with eq.~(4),
this causes the conductance to decrease quickly as $L$  increases,
approaching
that for the strongly localized case.
However, the $d=2$ case is more tricky: for high $G>>1$, the scaling function is so
close to zero, that minor correction may potentially change
the sign of $\beta$, therefore the situation requires a special attention.

\vspace{0.1in}
\underline{a) Spin-orbit scattering case} (Hikami, Larkin, Nagaoka (1980)).
The electron interference is destroyed by spin-orbit random scattering
and the weak localization ``back-scattering'' correction vanishes
\cite{hikami80,altshuler&aronov,lee_review}.
More over, the spin-orbit scattering induces
a positive correction (``forward scattering'') so that
the $\beta$-function becomes $\propto + 1/2G$ for high $G$,
as schematically shown in fig.~\ref{beta}\,b. Note, that this results
is valid only for a hypothetical 2D system of non-interacting particles.

\vspace{0.1in}
\underline{b) Interaction quantum corrections in the diffusive regime}.
Altshuler-Aronov corrections take into account
$e-e$ interactions  in the $T\tau \ll 1$ limit
\cite{altshuler&aronov,finkelstein}:
\begin{equation}
\delta\sigma =-\frac{e^2}{2\pi^2 \hbar}\left[1 +
3\left(1-\frac{\ln\left(1+F_0^a\right)}{F_0^a}\right)\right]\ln(\hbar/k_BT\tau),
\end{equation}
where $F_0^a$ is the Fermi-liquid interaction parameter, dependent on $r_s$
(see further for more details).
In the $F_0^a\rightarrow 0$ limit,
 the total quantum correction $\delta\sigma_{\rm loc}
+\delta\sigma_{so}+\delta\sigma_{ee}$
is usually negative (backscattering). However, when
$F_0^a<0$ and $|F_0^a|$ is large, the total quantum correction may
lead to delocalization (``forward scattering'').  This case will be discussed
in Sec.~\ref{sec:e-e}
in more details.

\vspace{0.1in}
\underline{c) Interaction quantum corrections to the transport in the ballistic regime}
 $T\tau \gg 1$
have been calculated
in Refs. \cite{ZNA,gornyi}; they are discussed in
Sec.~\ref{sec:e-e}.

\section{An apparent MIT in 2D}
\begin{figure}
\begin{center}
\includegraphics[width=6.0in]{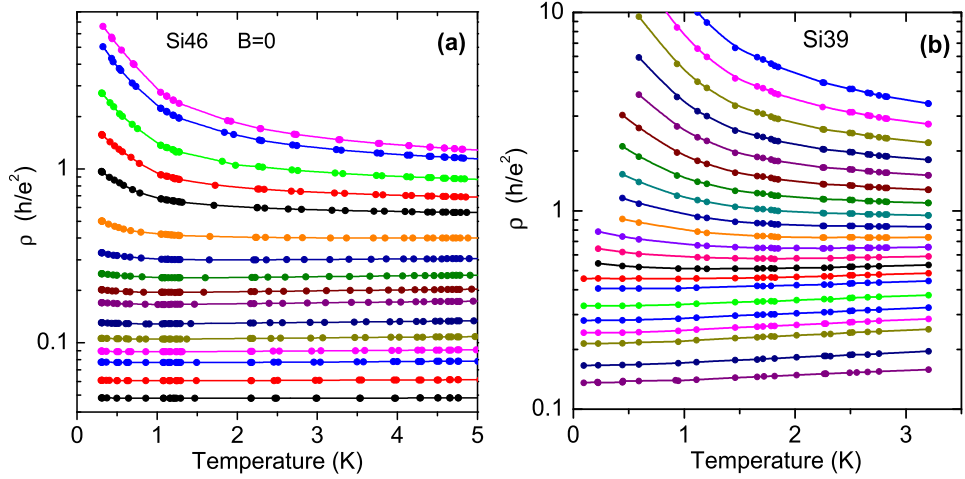}
\caption{Resistivity vs temperature for
disordered Si-MOS samples: (a) with peak mobility $\mu=1500$cm$^2$/Vs,
(b) $\mu=5100$cm$^2$/Vs.
The densities (in units of $10^{11}$\,cm$^{-2}$) on the left panel
 span  from 3.85 to 37.0 (from top to bottom),
on the right panel -- from 1.95 to 5.44.}
\label{Si46_Si39}
\end{center}
\vspace{-0.3in}
\end{figure}

A great body of experimental data accumulated since 1979
support the conclusions of the scaling theory \cite{bishop,uren}.
An example is shown in fig.~\ref{Si46_Si39}\,a.
The uppermost curves in figs.~\ref{Si46_Si39} show that
the resistivity for a very disordered (low-mobility) Si-MOS sample at low densities
 exhibits an exponentially strong increase with cooling.
This is consistent with motion to the left hand side along the
logarithmic part of the $\beta$-function in fig.~\ref{beta}\,a.
For higher densities (lower curves in fig.~\ref{Si46_Si39}),
the exponential increase of the resistivity is replaced by
a weak temperature dependence. The latter one
at very high carrier densities in Si-MOS  or in GaAs samples
is logarithmic   (see, e.g. fig.~\ref{Gmax}) and consistent
 with the theory ($d\rho/dT >0$).

As sample mobility increases, and, correspondingly,
the density of the crossover (from the exponential to logarithmic $T$-dependence)
decreases \cite{highmu}, the temperature derivative $d\rho/dT$ obtains
``wrong'' positive sign
(compare figs.~\ref{Si46_Si39}\,a, b, and fig.~\ref{high_mu}). This effect was noticed
more than two decades ago \cite{cham80,smith86,dorozh88}
and discussed in terms of the temperature-dependent
screening \cite{stern80,das86,gd86}.
As will be shown below, the sign of $d\rho/dT$
is determined by the
interaction parameter $F_0^a$, which in its turn, depends primarily on the electron density.

The strong indication that the prediction of insulating states in
2D may not be universally valid was obtained in 90s in studies of
the quantum Hall liquid to insulator transitions
 \cite{reentrant_dio,dioPRB92,hall93}.
Extended states, which in high magnetic field $B$ are located at
centers of Landau bands, were found experimentally to merge and
remain in a finite energy range as $B \rightarrow 0$. This
behavior is not expected within the scaling theory, which predicts
that the extended states ``float up'' in energy in that limit
\cite{halperin,khmelnitskii,laughlin}. A serious challenge to the
scenario of the insulating ground state in 2D arose in 1994 and
the years following, when metallic behavior of electrical
transport in zero magnetic field  was observed in  high mobility
Si MOSFET structures \cite{PRB94,PRB95} (see fig.~\ref{high_mu}).

Subsequently, qualitatively similar metallic states were found in
other high mobility 2D systems (for an extensive bibliography,
see Ref.~\cite{sk_RMP}). These include $p-$ and $n-$Si/SiGe--
\cite{coleridge,lam97,Brunthaler}, $p$-GaAs--
\cite{pepper,shahar,papadakis,mills,noh03}, $n$-GaAs--
\cite{ngaas,ensslin}, $n-$AlAs--heterostructures  \cite{alas},
inverted Si-on-insulator structures \cite{SOI} and back-gate
biased Si-MOS structures \cite{fowler97}. The most important
findings of these studies can be summarized as follows:\\

\underline{Features  in zero magnetic field}:
\begin{itemize}
\item If the density of 2D carriers $n$ is larger than a sample-dependent
value $n_c$, the $\rho(T)$ dependence is metallic-like, i.e. $d\rho/dT>0$
(see fig.~\ref{high_mu}).
The
changes in $\rho$  can be almost an order of magnitude
(in high mobility Si-MOSFETs  \cite{PRB94}).

\item
The strong metallic-like $\rho(T)$-dependence is characteristic
of a wide range of densities
rather than of the critical regime; it sets in
for relatively high temperatures $T \leq T_F$
in the ballistic rather than diffusive regime \cite{weakloc} (see fig.~\ref{Domains_Si43}).

\begin{figure}[htb]
\begin{minipage}[t]{.48\textwidth}
\includegraphics[width=2.5in]{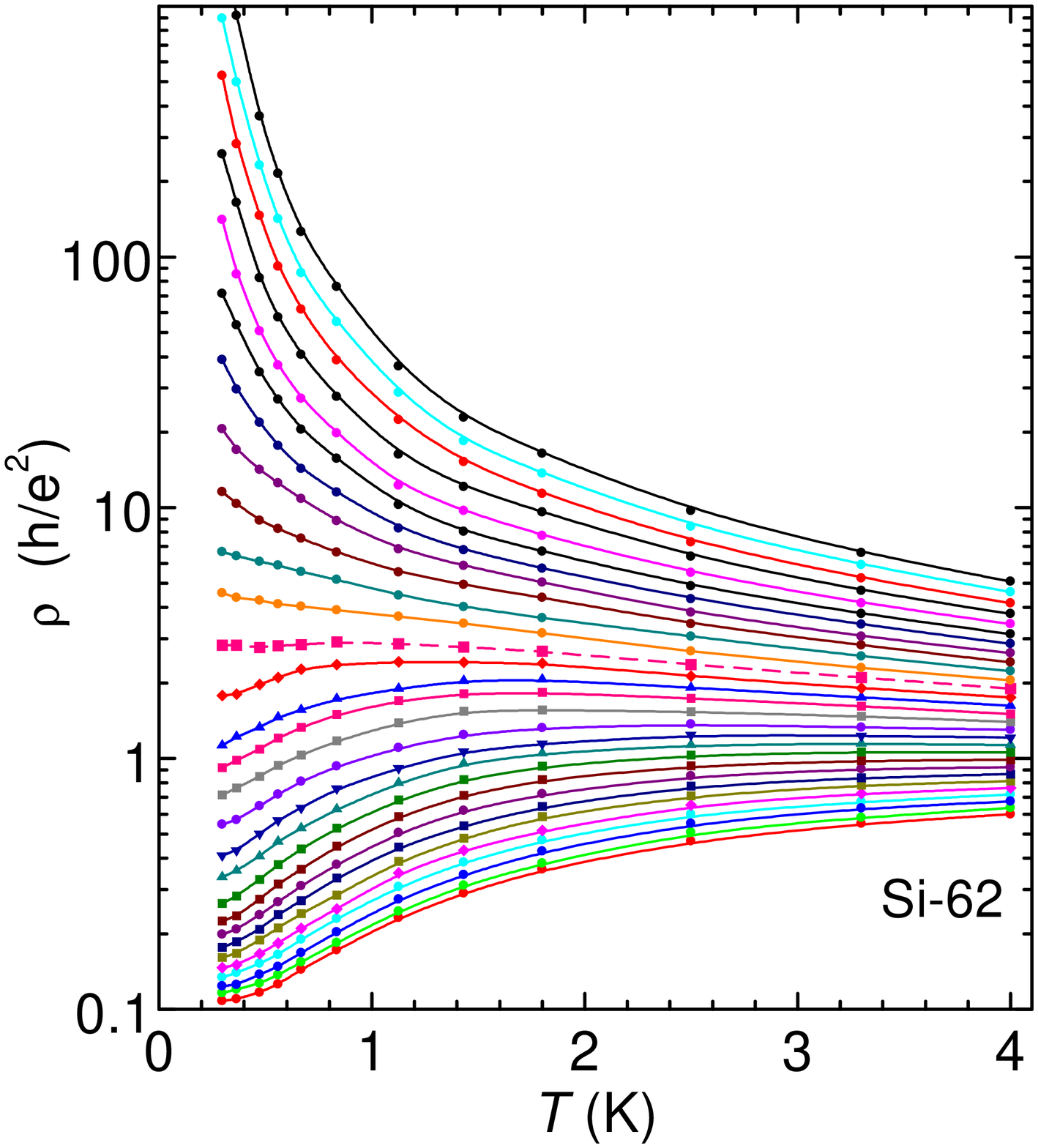}
\caption{Resistivity vs temperature for HIGH mobility  Si-MOS
sample \protect\cite{JETPL98scaling} in the critical regime
$n\approx n_c$. Electron density decreases from 1.326 in steps of
0.0218 (from  bottom to top) and in unites of $10^{11}$cm$^{-2}$.}
\label{high_mu}
\end{minipage}
\hfil
\begin{minipage}[t]{.47\textwidth}
\includegraphics[width=2.4in]{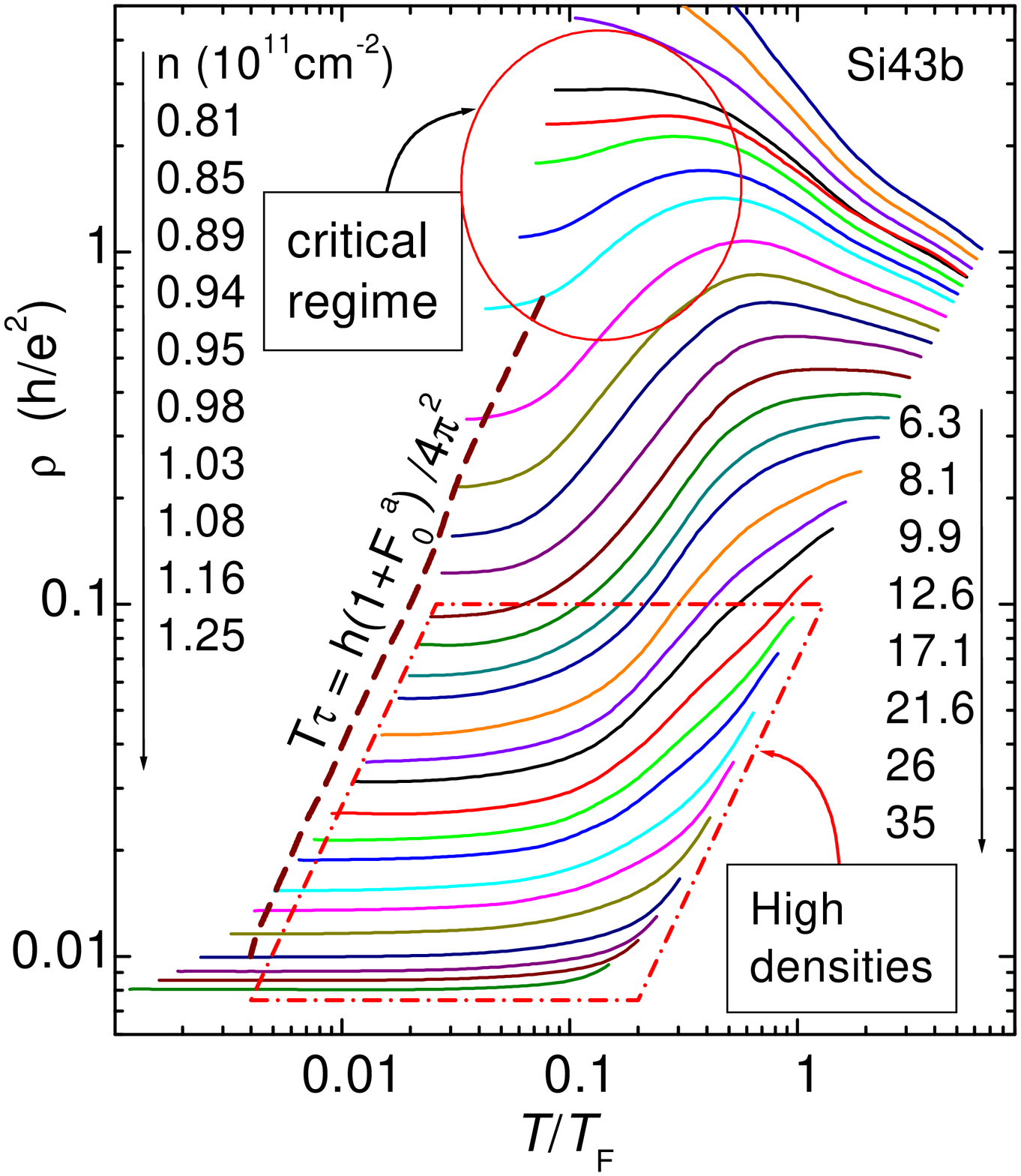}
\caption{Resistivity vs $T/E_F$ in a wide range of densities.
The diffusive/ballistic border for $T\tau$ \protect\cite{ZNA}
is calculated based on the data \protect\cite{weakloc} with
renormalized $m^*$.}
\label{Domains_Si43}
\end{minipage}
\end{figure}

\item When $n<n_c$, both the sign ($d\rho/dT<0$) and the functional (exponential)
form of the $\rho(T)$ dependence are characteristic of the insulating behavior
(see figs.~\ref{Si46_Si39},\ref{high_mu}).
\item At much higher densities ($n>30\times n_c$), the derivative $d\rho(T)/dT$  is
metallic (positive)
at higher temperatures $T\sim T_F$, but $\rho$ exhibits an slow (logarithmic)
``insulating'' upturn at lower $T$, consistent with
eq.~(2)
\cite{Gmax,hamilton99,simmons99} - see fig.~\ref{Gmax}.
Qualitatively, conduction
at high density is consistent with the weak localization picture.

\item
The $\log(R(T))$ data show a reflection symmetry with respect to
$n_c$ \cite{PRB94,PRB95} (see fig.~\ref{high_mu}), with  exclusion of the low-temperature data,
which in the $T\rightarrow 0$ limit diverge for $n<n_c$, but do not fall to
zero for $n>n_c$.
This  signifies that at intermediate temperatures, the $\rho(T)$-dependence
is empirically represented by a logarithmic function $\rho(T) \propto
\exp\left[\pm(T_0/T)^p\right]$ with $p \sim 1 $ and the sign equal to that of  $(n_c -n)$
\cite{JETPL98scaling};
such dependence obviously fits the one-parameter scaling function in the vicinity
of a critical point  shown in fig.~\ref{beta}\,b.
\item
For high-mobility samples, the resistances of the metallic and
insulating phases, measured as a functions of $T$ and $n$,
exhibit scaling properties with respect to $T/T_0$
\cite{PRB94,PRB95,JETPL98scaling} demonstrated in
fig.~\ref{scaling}. The energy scale $T_0$ goes to zero on both
sides of the transition; such behaviors are reminiscent of a
critical point of a continuous quantum phase transition
\cite{vlad_97}. This interpretation, however,
seems to work only for intermediate temperatures
 and fails in the $T\rightarrow 0$ limit. For a more detailed review
and phenomenological discussion of the  experimental data in the critical regime,
see ref.~\cite{akk}.
\begin{figure}[htb]
\begin{minipage}[t]{.48\textwidth}
\includegraphics[width=2.4in]{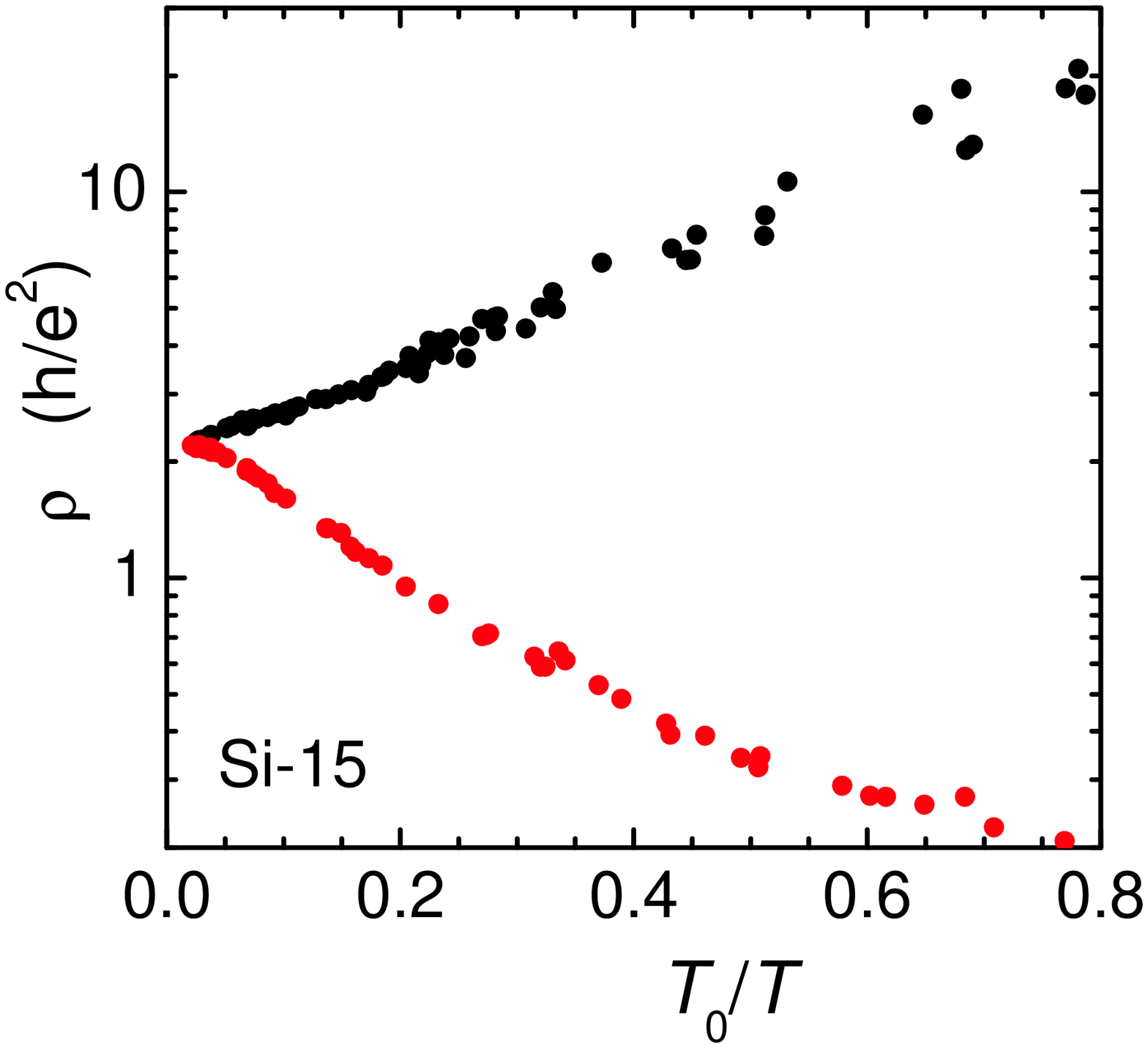}
\caption{Resistivity for high-$\mu$ sample
scaled vs $T_0/T$ \protect\cite{JETPL98scaling}}
\label{scaling}
\end{minipage}
\hfil
\begin{minipage}[t]{.48\textwidth}
\includegraphics[width=2.4in]{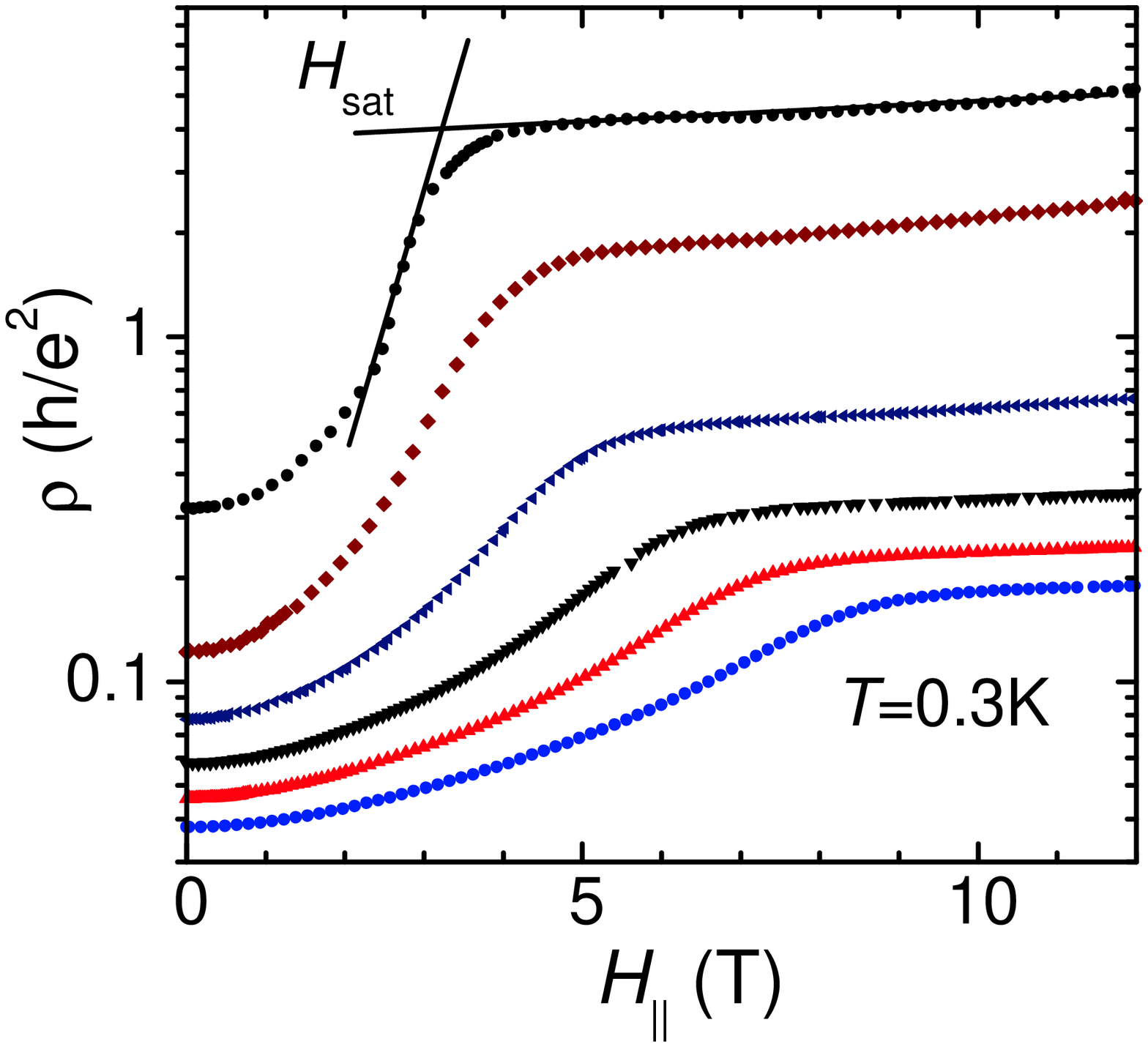}
\caption{$R(H_\parallel)$
dependences for different densities \protect\cite{anyso}.}
\label{R(H)}
\end{minipage}
\end{figure}

\vspace{0.1in}
\underline{Features in the in-plane field}
\item In Si-MOSFET and p-GaAs structures, $\rho$ increases strongly
with  $H_\parallel$ \cite{pudalov97,breakdown98,suppression97,okamoto99,yoon99,anyso,eng02}
and shows saturation at a field, which is approximately equal to that for
complete spin polarization \cite{vitkalov_doubling}; this is
most sharply pronounced in Si-MOS structures
\cite{suppression97,breakdown98,pudalov97,anyso}.
\item
As $B_\parallel$ couples to spins rather than to orbital motion,
it follows that the spin-related mechanism of interactions
(i.e. exchange) is of major importance. More specifically,
when $H_\parallel$ field switches off the spin-degrees of freedom,
the $R(T)$ dependence flattens and MIT disappears \cite{vitkalov_01,disorder}.
\end{itemize}

\vspace{0.1in} As $n_c$ is quite small ($\sim 10^{10}-10^{11}$
cm$^{-2}$), the corresponding $r_s$ values are large, an order of
10; therefore, it is reasonable to assume the $e-e$ interaction
to be one of the major driving forces in the above listed phenomena. In
the following section, the $e-e$ interaction effects will be
discussed in more detail. We shall show, in particular, how the
listed features find a natural explanation within the frameworks
of the quantum interaction corrections at intermediate
temperatures.

\section{Quantitative studies of the electron-electron interactions}
\label{sec:e-e}
\underline{4.1. Fermi-liquid renormalization of electron parameters in 2D systems}\vspace{0.05in}\\
The Landau Fermi-liquid (FL) theory \cite{landau,pines}
introduces a number of interaction parameters
to describe the interacting system in terms of the
renormalized $g^*$-factor Land\'{e},
effective mass $m^*$ etc.:
\begin{equation}
\label{eq:FL-renormalization}
F_0^a = \frac{2}{g^*}-1, \qquad  F_1^s =2\left(\frac{m^*}{m_b} - 1\right).
\end{equation}

\begin{figure}
\begin{center}
\includegraphics[width=1.0\textwidth]{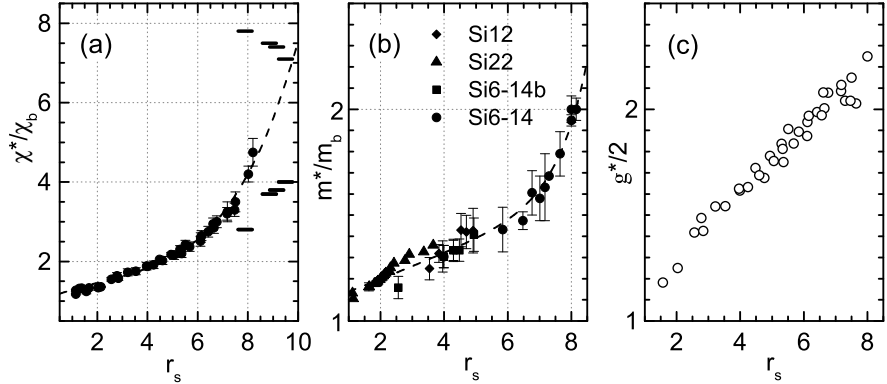}
\caption{Renormalized spin susceptibility (a), effective
mass (b), and $g^*$-factor (c) measured with Si-MOS
samples\protect\cite{gm,granada03}.} \label{chi_g_m}
\end{center}
\end{figure}

Recently, these parameters have been determined
experimentally in a wide range of densities, in measurements of the
quantum oscillations of conductivity (Shubnikov-de Haas (SdH) effect)
in tilted or crossed weak magnetic fields
\cite{okamoto99,gm,crossed,zhu}. In these experiments, the renormalized spin susceptibility
$\chi^*$ is obtained with no assumptions:
\begin{equation}
\label{eq:chi*}
\chi^*= g^*g_b m^* \mu_B \frac{e}{h}, \qquad \frac{\chi^*}{\chi_b} = \frac{g^*m^*}{2m_b}=
\frac{\left(F_1^s+2\right)}{\left(F_0^a +1\right)},
\end{equation}
where $g_b$, $m_b$, and $\chi_b$ are the corresponding band values.
Figure~\ref{chi_g_m} shows the $r_s$-dependence of the
measured renormalized $\chi^*$ and $m^*$-values
\cite{okamoto99,gm}; the renormalized $g^*$-values are obtained
from these two sets of data as $\chi^*/m^*$.
Other experimental approaches consist in scaling the
$\sigma(H/H_0)$-data for different densities
\cite{vitkalov_scalingMR_01,vitkalov_scalingMR_02,shashkin_R(B)scaling},
or in fitting the  $\sigma(T)$ and $\sigma(H_\parallel)$
dependences  to the  calculated quantum corrections \cite{ZNA}, using
$F_0^a$ (and in some cases, also $m^*$)  as fitting parameter
\cite{shashkin_fitting_s(T)02,aleiner,prosk_PRL02,prosk_JPhysA03}.

The resulting $F_0^a$ values obtained for 2D electrons in GaAs
\cite{zhu} and in Si-MOS samples
\cite{shashkin_R(B)scaling,vitkalov_scalingMR_02,gm,aleiner,prosk_JPhysA03,kvon_111Si_2002,shashkin_fitting_s(T)02}
qualitatively agree with each other \cite{granada03} but differ
substantially from the data for 2D holes in GaAs
\cite{prosk_PRL02}, and, especially, from the data taken for low
carrier density \cite{noh03}; the reason for this discrepancy is
not clear yet \cite{granada03}.

Figure~\ref{chi_g_m} shows a strong increase in $\chi^*$, $m^*$,
and $g^*$ with $r_s$, which should significantly affect transport
at low densities. The growth in $\chi^*$ might reflect a tendency
to either ferromagnetic or antiferromagnetic transition; besides,
within the FL-description eq.~(\ref{eq:FL-renormalization}), the
$g$-factor should diverge as $F_0^a \rightarrow -1$. Therefore, a
separate,  interesting question is whether or not $\chi^*$, $m^*$
and $g^*$ diverge as $r_s$ increases, and whether the divergency
(if any) occurs at the MIT. Numerical calculations can not give an
unambiguous answer \cite{senatore,paola02,ceperley01} for the
single-valley system, whereas the 2D valley system is expected to
be more stable and to remain unpolarized up to the WC phase.

As a result of the renormalization, the energy scale $E_F$ diminishes.
However, in the vicinity of the transition (i.e. at the sample-dependent density $n=n_c$),
$m^*$ does not diverge \cite{polariz,granada03}; this may be easily seen,
e.g., from non-vanishing amplitude of the SdH oscillations measured at $T=0.03$\,K
for $r_s$ values up to 9.5
\cite{polariz}. Thus, the MI-transition is  not caused by simply
vanishing the $E_F$ value and the quasiparticles. In
Refs.~\cite{shashkin_R(B)scaling,vitkalov_scalingMR_01} the
in-plane magnetoresistance $\rho(H_\parallel)$ data have been
discussed in terms of a developing ferromagnetic instability at or
very close to $n_c$. However, direct measurements of the (a)
magnetization of 2D electrons \cite{reznikov}, (b)  period,  and
(c) sign of the SdH oscillations \cite{polariz,granada03} do not
confirm such possibility: the spin susceptibility remains finite
at $n \geq 7.7\times 10^{10}$cm$^{-2}$, the density which  is
lower than the critical $n_c$-value  for many samples.

The strong enhancement of the weak field spin susceptibility was predicted also
by the renorm-group (RG) theory \cite{finkelstein,castellani_chi_84,castellani_chi_86};
however the RG-theory is based on consideration of the diffusive modes of electron-electron
interaction in the regime $T\tau \ll 1$,
whereas the renormalized parameters (Fig.~10) are measured in
the ballistic interaction regime $T\tau >1$ (for a more detailed discussion, see Section 4.3).

\vspace{0.1in}
\underline{4.2. Implementation of the measured FL parameters to  the metallic-like transport.}
\vspace{0.05in}\\
A considerable progress has been also achieved  in the theory of quantum transport:
in Refs.~\cite{ZNA,gornyi}
the interaction corrections
to the conductivity have been calculated beyond the diffusive regime, in terms of
the FL interaction parameters.
In this section, we make
a  comparison of the data with the theory
and conclude that the ``metallic'' drop of conductivity with cooling in the regime
$\sigma \gg e^2/h$ can be accounted for by the interaction effects in electron ``liquid''
at temperatures which correspond to the ballistic regime $T\tau >1$.
This observation suggests that the anomalous
``metallic'' conduction in 2D, at least for densities
not too close to the critical density, is
the finite-temperature phenomenon rather than the
signature of a new quantum ground state.

The theory  \cite{ZNA}
considers backscattering of electrons at the scattering centers
and at the Friedel oscillations of the density of surrounding electrons.
The interference
between these scattering processes gives rise
to the quantum corrections, which are calculated to higher orders
in the interactions and leading order to the temperature.
The interference
gives rise to quantum corrections to the Drude conductivity
(in units of $e^2/\pi \hbar$):
\begin{eqnarray}
\label{correction}
& \sigma(T)-\sigma_D = \delta \sigma_C(T) + 15\delta\sigma_T(T) 
. &
\end{eqnarray}
\begin{eqnarray}
{\rm Here}\qquad \qquad \qquad \delta\sigma_C = x\left[1-\frac{3}{8}f(x)\right] -
\frac{1}{2\pi}\ln\left(\frac{E_F}{T}\right) \qquad{\rm  and} \nonumber\\
\delta\sigma_T  = A(F_0^a)x\left[1-\frac{3}{8}t\left(x,F_0^a\right)\right]-
B(F_0^a)\frac{1}{2\pi}\ln\left(\frac{E_F}{T}\right)
\end{eqnarray}
are the
interaction contributions in the ,,charge'' (exchange term and singlet part of the correlation terms)
and triplet channels,
respectively; $x=T\tau k_B/\hbar$,\hspace{0.2in}
$A(F_0^a)=F_0^a/(1+F_0^a)$, and $B(F_0^a) = 1-\ln(1+F_0^a)/F_0^a$.
 The prefactor 15 to $\delta\sigma_T$ reflects
enhanced number of triplet components:
two valleys and two spins produce 4 pseudospin components; interaction of two particles
involves $4\times 4 =16$ channels, one of them being singlet.

\vspace{0.1in}
\underline{Diffusive regime,  $T\tau \ll 1$}.
For 
large $|F_0^a|$ and $F_0^a<0$, $\delta\sigma(T)$ becomes positive.
E.g., for $F_0^a = -0.3$ (i.e. $g^* \approx 3$)
\begin{equation}
\delta\sigma = \frac{1}{2\pi}\left(1-15\times 0.2 +1 \right)\ln T = -\frac{1}{2\pi}\ln T.
\end{equation}
Thus, due to valley degeneracy, the quantum correction is constructive
and produces a ``forward scattering''.

\vspace{0.1in}
\underline{Ballistic regime, $T\tau >1$}.
The quantum correction $\delta\sigma(T)$ is quasilinear in $T$:
\begin{equation}
\delta\sigma(T) \approx \left[1+15 \frac{F_0^a}{1+F_0^a}\right](T\tau)
\end{equation}

\vspace{0.1in}
\underline{Comparison of the theory with experiment at high $G\gg e^2/h$ (see Fig.~\ref{Domains_Si43})}.
\begin{figure}
\begin{center}
\includegraphics[width=3.2in]{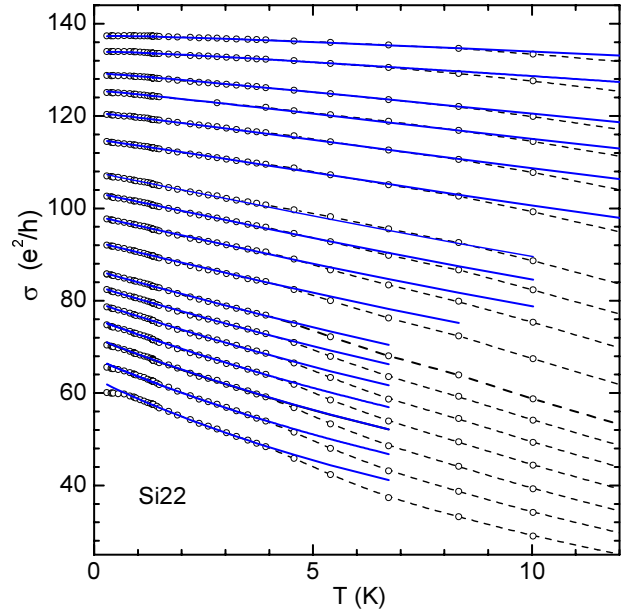}
\end{center}
\vspace{-0.2in}
\caption{Conductivity for sample Si22 vs
temperature \protect\cite{loc02}. The electron densities, from top to bottom are: $n=$
5.7, 6.3, 6.9, 7.5, 8.1, 8.7, 9.3, 10.5, 11.7, 12.9, 14.1, 16.5,
18.9, 21.3, 23.7, 28.5, 35.7 (in units of $10^{11}$cm$^{-2}$).
Dots represent the data, solid lines - the
theoretical curves with $F_0^a$ and $m^*$ from
Ref.~\protect\cite{gm}.}
\label{s(T)_comparison}
\vspace{-0.25in}
\end{figure}

The terms in eq.~(\ref{correction}) are functions of $T\tau/\hbar$  and
$F_0^a$.
The momentum relaxation time $\tau$
may be found from the Drude resistivity $\rho_D\equiv
\sigma_D^{-1}$ using the renormalized effective mass $m^*$
determined in Ref. \cite{gm}.
Thus, the comparison shown in fig.~\ref{s(T)_comparison}
has no fitting parameters,  beyond $\sigma_D$.
The determination of the latter value is transparent
\cite{aleiner}, it is obtained
by a quasilinear extrapolation of the experimental
$\sigma(T)$ data to $T=0$, by setting the logarithmic (diffusive) terms to zero.
Almost entire temperature range in fig.~\ref{s(T)_comparison} belongs to the ballistic regime.
In view of the quantitative agreement of the experimental data with theory,
we conclude that the initial slope of the $\sigma(T)$
dependence is well described by the quantum interaction corrections
eqs.~(\ref{correction})
with $F_0^a(r_s)$ and $m^*(r_s)$ values
determined independently (see fig.~\ref{chi_g_m}).
Several experimental teams came
to a similar conclusion
for $p$-GaAs/AlGaAs \cite{prosk_JPhysA03,noh03} and
Si MOSFETs \cite{shashkin_fitting_s(T)02,kvon_111Si_2002,vitkalov_R(B)03,aleiner}.

\vspace{0.1in}
We conclude this section by listing the achieved results:
\begin{itemize}
\item
Not too close to the critical regime, $\sigma  \gg  e^2/h$, the ``metallic''
strong $T$-dependence  results from
the quantum  interactions at intermediate range of $T$'s.
\item
As $T$ decreases further,
 in the diffusive regime,  $\sigma(T)$ is expected to vary $\propto
\ln T$ with $d\sigma/dT <0$ for sufficiently large
$g^*$ (large $|F_0^a|$) and large number of valleys.
Then, for $T <T_0$
(determined by the intervalley  $h/\tau_{vv}$ and spin-spin scattering rate $h/\tau_{ss}$),
the enhancement factor (15) should vanish and the localizing terms are expected to overwhelm.
\item
If the  true metallic phase in 2D exists, it must
of a non-FL type \cite{vlad_97,chakravarty_98,chakra_wigner_glass}.
Several non-FL theoretical models
have been recently suggested, such as
phase separation model (Wigner solid inclusions in the FL)
\cite{spivak}, an electron-paired state with broken U(1) symmetry \cite{pp},
magnetic phase transition etc. However, so far no clear signatures of the
non-FL behavior have been detected experimentally.
\end{itemize}
\vspace{0.1in}
\underline{4.3. Transport in the critical regime, $\sigma \sim e^2/h$ and $n\approx n_c$}\\
In view of the encouraging comparison with the theory of
interaction corrections in the high density regime, it seems
attractive to extend the comparison to the critical regime, where
$\sigma \sim e^2/h$ (see fig.~\ref{Domains_Si43}). Quantum
corrections theory is inapplicable for large conductivity changes
$\delta\sigma/\sigma \sim 1$ and in the vicinity of the critical
point, where $\sigma \sim e^2/h$. The method now commonly in use
is a generalization of the nonlinear $\sigma$-model theory, which
has been developed by Finkel'stein \cite{finkelstein} and by
Castellani, Di Castro et al. \cite{castellani_chi_84,castellani_MIT_84}. The
renormalization-group  method is an extension of the same
diagrams of the interaction theory to the case of the strong
disorder and interaction \cite{belitz}. The RG equations describe
renormalization of both, disorder (i.e. $\rho$) and interaction
parameters
 as the length scale
varies with $T$ \cite{finkelstein,castellani_MIT_84,dicastro}:

\begin{eqnarray}
\label{RG} \frac{d\rho}{dy} & = &
\frac{\rho^2}{\pi}\left[1+1-3\left(\frac{1+\gamma_2}{\gamma_2}
\ln(1+\gamma_2)-1\right)\right]\nonumber\\
\frac{d\gamma_2}{d y} & = & \frac{\rho(1+\gamma_2)^2}{2\pi},
\end{eqnarray}
where $y=-\ln(T\tau)$, $\gamma_2$ is the
Fermi-liquid amplitude for large angle scattering
($\gamma_2 \rightarrow -F_0^a/2$ in the weak-coupling limit),
and $\rho$ is in units of $h/e^2$.
The 1st term in the square brackets is the weak localization contribution,
the 2nd term is the $e-e$ interaction correction in the singlet channel
\cite{altshuler&aronov,lee_review},
and the last term is the contribution from 3 triplet modes.
Due to the difference in symmetry
of the singlet and triplet wavefunctions, the singlet and triplet terms have different sign,
and favors localization and delocalization, correspondingly. For two-valley system, factor 3
should be again replaced by 15,
and the weak-localization term also becomes twice larger \cite{punnoose}.
\begin{figure}
\begin{center}
\includegraphics[width=5.0in]{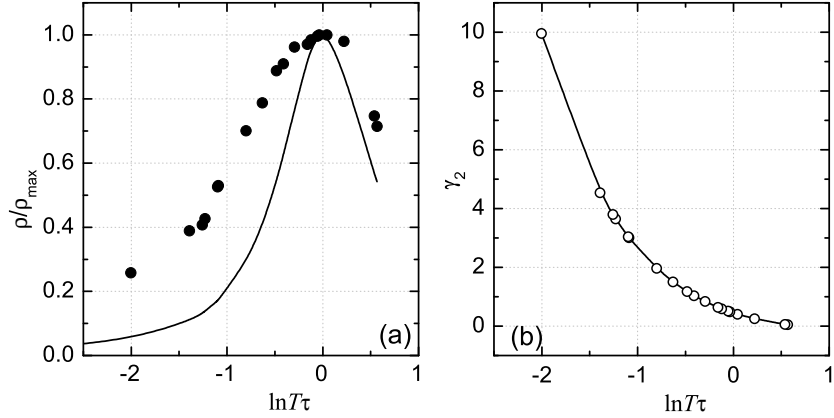}
\caption{a) Comparison of the $\rho(T)$ data normalized to its
maximum value ($1.35 h/e^2$) with solution of the RG equations (solid line)
eq.~(\protect\ref{RG}); b)
Renormalized  interaction parameter $\gamma_2$
\protect\cite{punnoose}, calculated from the 2nd Eq.(11) with
$\rho(T)$ data shown on the left panel. The ,,0''-point at the $\ln(T\tau)$ scale
corresponds to $T=1.8$K. Sample Si15, electron density $n=0.88\times 10^{11}$cm$^{-2}$.
}
\label{Finkelstein_Si15}
\end{center}
\vspace{-0.3in}
\end{figure}

The above equations have a universal solution, which has been 
compared with experimental
$\rho(T)$ data in Ref.~\cite{punnoose}.
We present in fig.~\ref{Finkelstein_Si15}
a similar comparison,
which 
confirms a certain similarity between the $\rho(T)$-data and the theory.
There is a limitation though, the
 eqs.~(\ref{RG}) are perturbative, i.e. are derived in the lowest order in $\rho$,
therefore, the  comparison with experimental data for $\rho(e^2/h) \sim 1$ can be only qualitative.

The application of the RG perturbative equations to the experimental data
is based on the following assumptions:\\
(i) the critical regime $n\approx n_c$ belongs to the diffusive
domain of {\it e-e}-interactions $T\tau \ll 1$;\\
(ii) the $\rho(T)$ maximum
signifies
a turning point from localization
to delocalization behavior due to the renormalization
of the interaction parameter (primarily, $\gamma_2$) with $T$;\\
(iii) the decrease in $\rho$ with cooling takes place mainly for the account
of the strong growth in the interaction parameter $\gamma_2$ (and hence, $F_0^a$).

\vspace{0.1in}
The first assumption does not agree well with the empirical diagram
in fig.~\ref{Domains_Si43}; therefore, its application would
require to re-interpret the definition of the diffusive/ballistic border,
i.e. to find a different interpretation
for the $\tau(n)$ or $F_0^a(n)$ data in the critical regime.
In Ref.~\cite{punnoose},
the assumption (ii)  was presumed to
be fulfilled in the most clean samples; however,  some
high mobility Si-MOS samples do not exhibit a pronounced $\rho(T)$ maximum
in the critical regime \cite{kk,JETPL98scaling} (see, e.g., fig.~\ref{high_mu}).
Furthermore, the $\rho(T)$
data in the critical regime, as a rule, show a strong
sample-dependent and non-universal behavior \cite{JETPL98scaling,cooldowns}.
It is possible though that the relevant disorder is related with
short-range fluctuating potential at the interface
which may favor inter-valley scattering and, thus reduce a
number of the triplet terms in eq.~(\ref{RG})
but is not significant for the mobility at low $T$.

Regarding assumption (iii), figure~\ref{Finkelstein_Si15}\,b shows that the interaction parameter
$\gamma_2$, determined from the experimental data, starts growing
at rather high temperatures,
$\sim 1.8$\,K for $n\approx n_c$; the latter behavior of $\gamma_2$
is necessary in the RG-theory to turn down the resistivity flow. Thus, the renormalization
of $\gamma_2$ should occur in a quite accessible range of temperatures
(0.3 - 1.8)K. However,
the anticipated strong $T$-dependence
of the  renormalized spin susceptibility, eq.~(\ref{eq:chi*}),
\cite{castellani_chi_84}
is not observed experimentally at these temperatures in SdH-effect measurements:
$\chi^*/\chi_b \propto g^*m^*/2m_b$ changes only within 2\%
in the above temperature range \cite{granada03}.

An alternative theoretical approach \cite{vlad_97} suggests that the critical
behavior in the vicinity of $n_c$ is dominated by the physics of
the insulating state, namely by the melting of the disordered
Wigner crystal \cite{chakra_wigner_glass}, electron glass
\cite{thakur99}, or an inhomogeneous state consisting of
inclusions of WC into the 2D liquid \cite{spivak}. Corresponding
theories has not reached yet a fully predictive stage. We note that
signatures of the glassy behavior were reported in Ref.~
\cite{snezana02}.

\vspace{0.05in}
\underline{4.4. Transport in the presence of the in-plane field}\\
Figure  \ref{R(H)} shows a typical  behavior of $\rho(B_\parallel)$.
At low densities $n\sim n_c$, as field increases, transport
becomes temperature activated (hopping regime)
\cite{disorder}.
When $n$ approaches $n_c$, by definition,
$n_c(H_\parallel)$ tends to
$n_c(H_\parallel =0) \equiv n_c$,  therefore, the range of magnetic fields
where the magnetoresistance (MR) may be studied in the regime of ``metallic'' non-activated conduction,
shrinks to zero.

\vspace{0.1in}
\underline{Regime of high densities $n\gg n_c$, $G\gg e^2/h$.}
According to the theory of interaction corrections \cite{ZNA},
\begin{equation}
\label{eq:s(B)}
\delta\sigma(H_\parallel)\equiv \sigma(0,T)-\sigma(E_Z,T)
\approx \frac{e^2}{\pi\hbar}f(F_0^a) \frac{T\tau}{\hbar} \left[ K_b\left(\frac{E_z}{2T},F_0^a\right) \right]
\end{equation}
where $K_b \approx (E_Z/2T)^2 f(F_0^a)$  in the low field  limit $E_Z/2T \ll 1$, and
$K_b \approx (E_Z/T)f_2(F_0^a)$  in the limit  $E_Z/2T \gg 1$ limit.
Correspondingly, as field increases, $\delta\sigma(H)$
should increase initially $\propto H^2/T$, and than  $\propto H$.

Experimental data, in general,  show similar behavior \cite{vitkalov_R(B)03,gm,aleiner}
with a transition from parabolic to linear dependence;
however, in contrast to
the temperature dependence of $\rho$,
the agreement with  theory \cite{ZNA} is only qualitative \cite{aleiner}.
In high fields, $g^*\mu H_\parallel \gg T$,
the magnetoresistance
deviates substantially from the theory,
and the deviations are sample dependent \cite{aleiner}.
As density decreases and field increases,
the deviations of the measured MR from the theory grow
and reach  a factor of two  at $n=1.5n_c$ \cite{aleiner,disorder}.
The
measured $\sigma(H,T)$
scales with temperature somewhat different from  eq.~(\ref{eq:s(B)}),
which predicts  $\delta\sigma \propto -(H^2/T)$ in low fields.

\vspace{0.05in}
\underline{Critical regime of densities $n\approx n_c$, and  $\sigma \approx e^2/h$.}
As mentioned above,
the magnetoresistance studies in the critical regime
are restricted to the fields vanishing to zero
as $n\rightarrow n_c$. Nevertheless, down to densities $n\approx 1.1n_c$.
the magnetotransport can be safely studied in the regime $g^*\mu H_\parallel \ll T$.
The experimental data in low fields scales
as $\delta\sigma \propto -(H^2/T^p)$,
where $p$  increases  from 1.1 to 1.6 as
density decreases from $5n_c$ to $1.2 n_c$.
Comparing this empirical scaling law
with that predicted by the RG theory
\cite{finkelstein,castellani_98},
\begin{equation}
\sigma(H,T)-\sigma(0,T) = -0.084\frac{e^2}{\pi h}\gamma_2\left(\gamma_2 +1\right)
\left(\frac{g\mu H}{k T}\right)^2,
\end{equation}
we conclude that the interaction
parameter $\gamma_2$ (roughly, $\propto |F_0^a|$) {\it decreases} as temperature decreases.
This result  is not consistent with
the main idea of the two-parameter scaling, where $\gamma_2$ is
expected to increase and $\chi^*$ diverges $\propto T^{-4/3}$
\cite{castellani_chi_84} as temperature decreases (i.e., the
length scale increases). It is important also that the direct SdH
measurements in low $H_\parallel$ fields do not confirm strong
$T$-dependence of $F_0^a$ \cite{granada03}. It might be that the
critical regime where the RG-description is applicable is  much
narrower, than the range of densities ($\approx 20\% n_c$) where
the $\rho(T)$ data exhibit a pronounced maximum. If this is the
case, the $\rho(H_\parallel,T)$ analysis should be restricted
also to much lower magnetic fields. We note that right at the
transition and for lower densities $n\leq n_c$, the $\rho(T)$
displays an approximately $T$-activated exponential dependence
$\rho(H,T) \propto \exp(\Delta /T)$ with $\Delta \propto
(n-n_c(H)) \propto H$ \cite{disorder}; the resulting dependence
$\exp(H/T)$ is not easy to distinguish from the
$(H/T)^2$-dependence.

In refs.~\cite{disorder,anyso,cooldowns,aleiner},
it was suggested that transport
in the critical regime is driven  not only by ``universal'' effects of interactions among
the itinerant electrons, but
also by  interactions of itinerant electrons with localized
ones, the latter issue (short-range aspect of interactions)
 is missing in most of the existing theories.

\vspace{0.05in}
\underline{Homogeneity.}
There is an important
issue, whether or not a   spatial inhomogeneity
develops in the 2D electron system as the carrier density decreases;  such
inhomogeneity was suggested to
cause the percolation-type MI-transition \cite{shi02,meir99}.
Though this issue was considered in a number of theoretical  papers, there is no clear
experimental evidence for such mechanism to be the major driving force in the experimentally observed
phenomena in high mobility samples.

\section{Summary}

I will present below a brief pedagogical summary of the issues which has been learnt
in the considered field, and those which still remain unanswered.
\begin{itemize}
\item
For high  densities and high conductivities,
in the ballistic regime $T\tau \gg 1$,
it is now widely accepted that
the metallic $T$-dependence of the conductivity
is a finite temperature  effect caused by $e-e$ interactions. However,
(i) in the diffusive regime $T\tau \ll 1$,  a thorough comparison with theory
is missing; some data reveal a substantial disagreement \cite{aleiner} with theory,
which may be attributed, e.g.,
to the intervalley scattering;
(ii)
a successful comparison with interaction theory has not been
demonstrated yet for the same set of $\rho(H, T)$-data for both zero and non-zero in-plane
field.
\item
For low densities, in the critical regime  $n \approx n_c$,
the  observed  one-parameter critical behavior
of conduction $\sigma(T)$ seems to be characteristic of
intermediate temperatures only. For much lower temperatures,
in the diffusive regime,
the potential critical behavior  remains unexplored.
\item
So far, all data find a reasonable explanation within the FL theory;
no clear signature of the non-FL behavior was detected.
\item
Both, the growth of the spin susceptibility as density decreases and the
non-monotonic $T$-dependence of resistivity
in the critical regime $n \approx n_c$, are in a qualitative agreement
with the RG theory. However, some other experimental  results are inconsistent
with the  theory; the discrepancy requires more detailed
analysis of the experimental data and the RG scenario of the MIT in 2D.
\item
 The behavior of the renormalized spin susceptibility and effective mass for 2D holes at
 high $r_s$ values remains puzzling; the reason of the  discrepancy
 between the 2D hole and electron systems is not clear.
\item
By now,
there is no complete theory which would include
short-range and high-energy aspects
of $e-e$ interactions, inter-valley scattering, etc.
\item
The interesting ideas of the potential spontaneous polarization transition
in the spin or valley system require investigations at substantially
higher $r_s$ values.
\item
The detailed experimental verification of the possible formation of local moments,
melting of the Wigner glass and other non-FL scenarios is currently lacking.

\end{itemize}

\acknowledgments
I am grateful to the Italian Physical Society for the hospitality,
and to G.~Giulliani and G.~Vignale for stimulating  fruitful atmosphere at
the 2003 Enrico Fermi school.
I  would like to thank
C.~Di Castro, D.~Ceperley, G.~Senatore, and B.~Tanatar for  interesting discussions.
The reviewed data were taken with  M.~Gershenson, H.~Kojima,
E.~Dizhur, G.~Brunthaler, A.~Prinz, and G.~Bauer, whom I would
like to specially thank. The high mobility Si-MOS samples were
manufactured  in collaboration with M.~Vernikov, L.~Pazinich, and
S.~G.~Semenchinskii. The research done by our experimental group
was supported by  INTAS, RFBR, and Russian grants from the
Ministry for science and education, and the Presidential program
of the support of leading scientific schools.


\begin{thebibliography}{0}
\bibitem{highmu}As sample  mobility increases,
the disorder and resistivity decrease
for the given carrier density. Correspondingly,
the onset of the exponential insulating behavior
is shifted to lower $n_c$ values,
thus resulting in
larger  values of $r_s$ and of the interaction parameter $|F_0^a|$.
\end{thebibliography}
\end{document}